\documentclass{raa_twocolumn}

\usepackage{graphicx,times}             
\usepackage{natbib}
\usepackage{amssymb,amsmath}
\bibpunct{(}{)}{;}{a}{}{,}
\usepackage[pagebackref=true]{hyperref}
\usepackage{comment}

\usepackage{caption}
\begin{document}

  \title{The Gamma-Ray Monitor onboard the SVOM satellite
}

   \volnopage{Vol.0 (202x) No.0, 000--000}      
   \setcounter{page}{1}          

  \author{Jian-Chao Sun\inst{1,*}\footnotetext{$*$Corresponding Authors, these authors contributed equally to this work.} \and 
  Yong-Wei Dong\inst{1,*} \and
  Jiang He\inst{1,*} \and
  Jiang-Tao Liu\inst{1} \and
  Lu Li\inst{1} \and 
  Rui-Jie Wang\inst{1} \and
  Xin Liu\inst{1} \and
  Li Zhang\inst{1} \and
  Min Gao\inst{1} \and
  Yue Huang\inst{1} \and
  Hao-Li Shi\inst{1} \and
  Li-Ming Song\inst{1,2} \and
  Wen-Jun Tan\inst{1,2} \and
  Chen-Wei Wang\inst{1,2} \and
  Jin Wang\inst{1} \and
  Jin-Zhou Wang\inst{1} \and
  Ping Wang\inst{1} \and
  Xing Wen\inst{1} \and
  Bo-Bing Wu\inst{1} \and
  Shao-Lin Xiong\inst{1} \and
  Juan Zhang\inst{1} \and
  Shuang-Nan Zhang\inst{1,2} \and
  Xiao-Yun Zhao\inst{1} \and
  Shi-Jie Zheng\inst{1}
   }
   

   \institute{State Key Laboratory of Particle Astrophysics, Institute of High Energy Physics, Chinese Academy of Sciences, Beijing 100049, China; {\it sunjc@ihep.ac.cn, dongyw@ihep.ac.cn, hejiang@ihep.ac.cn}\\
\and 
    University of Chinese Academy of Sciences, Chinese Academy of Sciences, Beijing 100049, China\\
\vs\no
   {\small Received 202x month day; accepted 202x month day}}

\abstract{The Gamma-Ray Monitor (GRM) is a key scientific payload onboard the Space-based Multi-band Variable Object Monitor (SVOM) satellite, designed specifically for the detection and study of gamma-ray bursts (GRBs). Launched into a 625~km low-Earth orbit on 22 June 2024, GRM serves as a large-area, wide-field-of-view instrument capable of observing the hard X-ray and soft gamma-ray emissions in the energy range of 15~keV to 5~MeV. Its primary scientific objectives include: promptly triggering and localizing GRBs (with particular sensitivity to short-hard GRBs), measuring spectral and temporal properties of bursts, monitoring charged particle fluxes in orbit. GRM successfully detected its first GRB (GRB~240627B) on 27 June 2024, and has since maintained a detection rate of more than 100 GRBs per year. Cross-instrument comparisons with detectors such as GECAM and \textit{Fermi}/GBM have validated the performance and data quality of GRM. This paper provides a comprehensive overview of GRM instrument design, reliability verification through ground testing, in-orbit triggering and localization algorithms, performance calibration, and preliminary in-orbit results, demonstrating its capability as a versatile gamma-ray all-sky monitor. 
}

   \authorrunning{J. C. Sun, Y. W. Dong and J. He et al. }            
   \titlerunning{GRM instrument overview}  

   \maketitle

%
%
\section{Introduction}           
\label{sect:intro}

Gamma-ray bursts (GRBs) are among the most energetic and enigmatic phenomena in the universe, originating from cataclysmic events such as massive star collapses or neutron star mergers. Since their serendipitous discovery in the 1960s, our understanding of these extreme events has been significantly advanced through observations by successive space missions. Pioneering missions such as BeppoSAX~\citep{boella1997bepposax}, Swift~\citep{gehrels2004The}, and Fermi~\citep{meegan2009The} have enabled groundbreaking discoveries in GRB localization and afterglow characterization. More recent contributions from missions such as HXMT~\citep{zhang2020overview} and GECAM~\citep{li2020the} have further enriched our understanding of GRB prompt emission mechanisms and their connection to multi-messenger astrophysics.

Despite these advances, fundamental questions of GRBs remain unanswered~\citep{2021RAA....21...55L,2021Galax...9...82G,2014IJMPD..2330002Z}, especially for short GRBs, such as their connection to gravitational wave events, and the extreme physical conditions in their environments. The SVOM (Space-based Multi-band Variable Object Monitor) satellite~\citep{Cordier+etal+2026a,Li+etal+2026,2022A&A...665A..40D,wei2020Progress,2020ApOpt..59.3049F,2019ExA....48..171M,wei2018Brief,2018SPIE10699E..21M,2017ExA....44..113B,Wei+etal+2016,2012SPIE.8443E..1OG,2011CRPhy..12..298P,2009AIPC.1133...25G}, a Sino-French cooperative mission, represents a significant advancement in GRB research. Launched in June 2024 into a 625~km low-Earth orbit, SVOM carries four scientific instruments covering gamma-ray, X-ray, and optical bands, enabling comprehensive studies of GRB prompt emission and afterglows.

The Gamma-Ray Monitor (GRM)~\citep{he2025svom,wen2021calibration,he2020bk,2013RAA....13.1381Z,dong2010SVOM} serves as a key payload onboard SVOM, designed as a large-area, wide-field detector operating in the 15--5000~keV energy range. GRM's primary scientific objectives include: triggering gamma-ray transients (including GRBs, SGRs, with particular emphasis on short-hard GRBs), measuring spectral and temporal properties of bursts, monitoring charged particle fluxes, and providing real-time trigger alerts for follow-up observations. Since its launch, GRM has demonstrated excellent performance, detecting numerous GRBs including the mission's first triggers GRB~240627B, GRB~240629A, and GRB~240702A~\citep{svom1stgcn}. Cross-instrument comparisons with GECAM and \textit{Fermi}/GBM have validated GRM's detection capabilities and data quality.

Within the SVOM payload suite, GRM works in synergy with the ECLAIRs telescope~\citep{Godet+etal+2026a}, the Microchannel X-ray Telescope (MXT)~\citep{Goetz+etal+2026}, and the Visible Telescope (VT)~\citep{Qiu+etal+2026} to form a complete multi‑wavelength observing system. ECLAIRs, a coded‑mask imager operating in the 4–250~keV band, provides accurate localization ($<10$~arcmin) of GRBs within its 2~sr field of view (FoV). GRM serves as the high‑energy sentinel of the SVOM payload suite, extending the energy coverage up to $\sim$5~MeV and enabling studies of the prompt emission, including jet physics and radiation mechanisms. Its large effective area and wide FoV ($\sim$2$\pi$~sr) make it highly sensitive to GRBs falling outside ECLAIRs' coded FoV. With its independent on‑board trigger algorithm—particularly effective for short, hard bursts—GRM can detect events that might otherwise be missed, thereby increasing the overall GRB detection rate of the mission. Once a burst is detected and coarsely localized by GRM, or precisely localized by ECLAIRs, MXT and VT perform follow‑up observations in X‑ray and optical bands to study the afterglow and measure its redshift. This multi‑instrument synergy is key to fulfilling SVOM's science goals, especially for multi‑messenger follow‑up of gravitational wave sources.

This paper provides a comprehensive overview of the GRM instrument, detailing its scientific objectives, instrumental design, ground calibration, preliminary in-orbit performance, and its role in advancing GRB research within the multi-messenger astronomy context.


\section{Overview of GRM Mission}
\label{sect:mission}

GRM is designed to significantly advance our understanding of GRBs and related high-energy astrophysical phenomena. Its scientific objectives encompass three primary domains: GRB physics investigation, including progenitor systems, jet mechanisms, energy dissipation processes, and radiation mechanisms; multi-messenger astronomy through complementary studies of gravitational waves, neutrinos, and high-energy cosmic rays; and cosmology and fundamental physics by probing early universe properties and testing physical laws under extreme conditions. Apart from GRB sciences, GRM can also detect other soft gamma-ray transients, including soft gamma-ray repeater (SGR)~\citep{2025ApJ...993...24W,2025ApJS..277....5X}, Solar Flare (SFL)~\citep{2021ApJ...918...42Z,2023SCPMA..6659611Z}, etc. Additionally, GRM's unique capabilities in low Earth orbit, combined with its wide FoV and broad energy coverage extending to several MeV, also make it ideally suited for studying Terrestrial Gamma-ray Flashes (TGFs)~\citep{2025SCPMA..6881011Y,2025SCPMA..6851013Y,2023GeoRL..5002325Z}, establishing it as a versatile instrument for both astrophysical and terrestrial high-energy phenomena.

The scientific objectives of GRM translate into the following specific requirements (SR):

\textbf{[SR1]} Detect diverse GRB populations including short GRBs (5~ms--2~s), long GRBs (up to 1000~s), and X-ray rich GRBs, as well as those outside ECLAIRs' FoV but within GRM's detection range, thereby enhancing ECLAIRs' GRB discovery capability.

\textbf{[SR2]} Provide rapid triggering for short GRBs, leveraging GRM's sensitivity to detect events that may not trigger ECLAIRs, similar to Fermi/GBM's enhanced short GRB detection compared to Swift/BAT.

\textbf{[SR3]} Observe GRBs across a broad energy band (15--5000~keV) from T0-5~min to T0+10~min, enabling comprehensive studies of temporal and spectral features including precursors, extended emission, and spectral components, with energy band overlap facilitating cross-calibration with ECLAIRs.

\textbf{[SR4]} Support synergistic observations of gravitational wave sources, targeting NS-NS and BH-NS mergers as established progenitors of short GRBs, building on the confirmed association between GRB 170817A and GW 170817.

\textbf{[SR5]} Detect and characterize TGFs with durations of 50~$\mu$s to several milliseconds across 10~keV to MeV energies, including potential electron-positron pair observations.

These science requirements lead to the following functional specifications (FR) for GRM:

\textbf{[FR1]} Provide spectral observations of GRBs from 15 to 5000~keV.

\textbf{[FR2]} Implement on-board triggering and duration measurement capabilities, with particular emphasis on short, hard GRBs.

\textbf{[FR3]} Measure GRB peak energies in near real-time across hard X-ray and soft gamma-ray bands, with combined analysis capabilities using ECLAIRs data for improved accuracy.

\textbf{[FR4]} Monitor and provide alerts for high particle flux regions, particularly the South Atlantic Anomaly (SAA).

\textbf{[FR5]} Deliver coarse localization information for detected GRBs.

\begin{figure}[h] 
   \centering
   \includegraphics[width=8.0cm, angle=0]{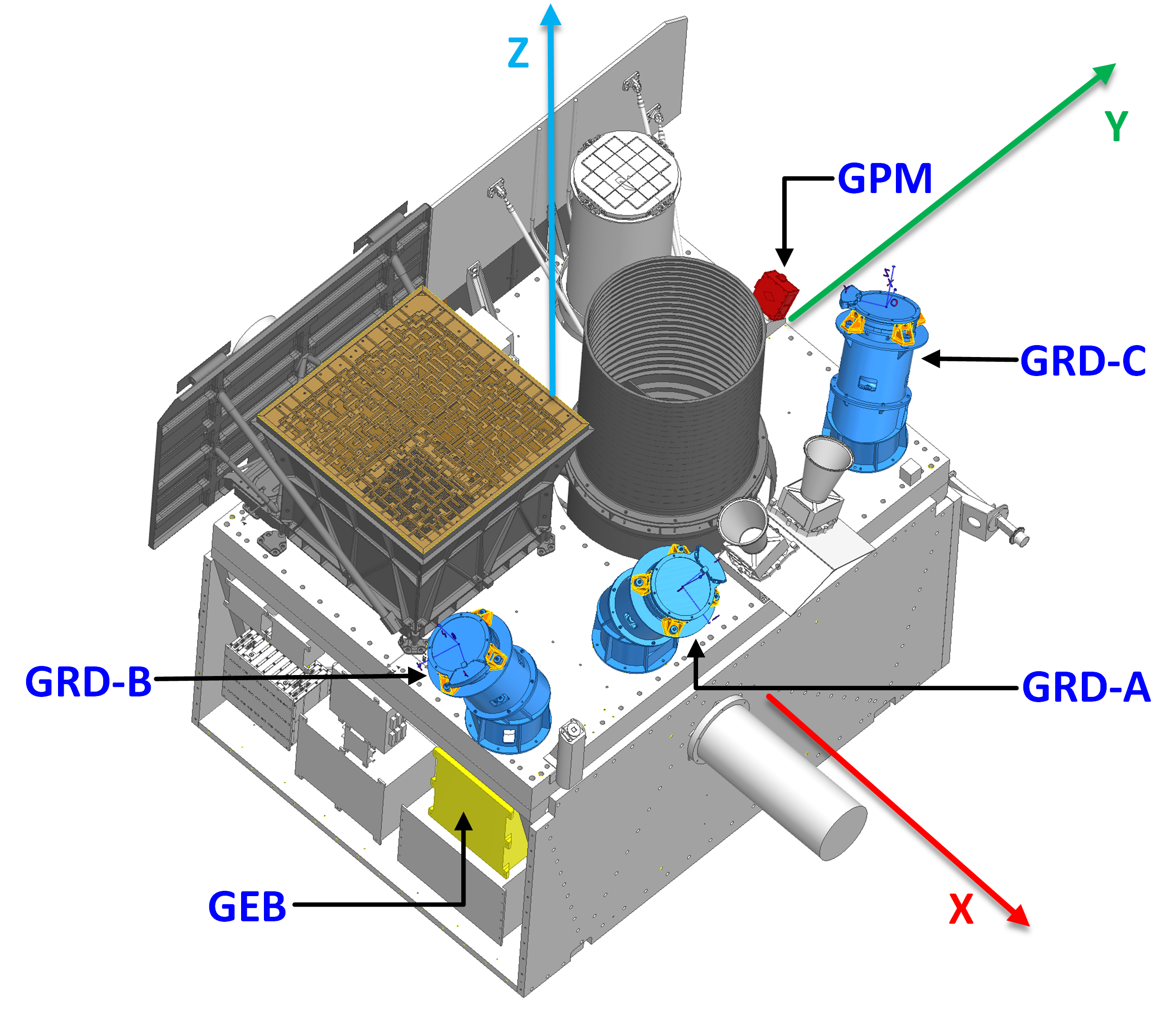}
   \caption{GRM instrument configuration onboard the SVOM satellite, showing the four primary payloads: GRM, ECLAIRs, MXT and VT. The GRM system comprises GRD-A/GRD01, GRD-B/GRD02, GRD-C/GRD03, GPM, and GEB units.} 
   \label{Fig1}
\end{figure}

\begin{table*}
\begin{center}
\caption{GRM performance specifications.}\label{Tab1}
 \begin{tabular}{llll}
  \hline\noalign{\smallskip}
\textbf{Performance Metrics} & \textbf{Required value} & \textbf{Actual value} & \textbf{Assessment method}    \\
  \hline\noalign{\smallskip}
    \begin{minipage}[t]{0.22\textwidth} 
    \  \\
    Energy range \\
    \  \\
    \end{minipage} 
    &
    \begin{minipage}[t]{0.22\textwidth} 
    \ \\
    15--5000~keV \\
    \ \\
    \end{minipage} 
    &
    \begin{minipage}[t]{0.22\textwidth} 
    GRD-A: 8--5927~keV\\
    GRD-B: 8--5675~keV\\
    GRD-C: 8--5638~keV\\
    \end{minipage}
    & 
    \begin{minipage}[t]{0.22\textwidth} 
    \ \\
    Measurement $\&$ calculation\\
    \ \\
    \end{minipage} \\ 
    Field of view & $\pm$60~degrees & $\pm$90~degrees & Measurement \\
    \begin{minipage}[t]{0.22\textwidth} 
    \ \\
    Sensitive area \\
    \ \\
    \end{minipage}
    &
    \begin{minipage}[t]{0.22\textwidth} 
    \ \\
    $>$200~cm\textsuperscript{2} (each unit)\\
    \ \\
    \end{minipage}
    &
    \begin{minipage}[t]{0.22\textwidth} 
    GRD-A: 201.31~cm\textsuperscript{2}\\
    GRD-B: 201.06~cm\textsuperscript{2}\\
    GRD-C: 201.31~cm\textsuperscript{2}\\
    \end{minipage}
    & 
    \begin{minipage}[t]{0.22\textwidth} 
    \ \\
    Measurement $\&$ calculation\\
    \ \\
    \end{minipage} \\
    Dead time & $<$8~\textmu s & $\sim$ 4.167~\textmu s & Measurement \\
    Temporal resolution & $<$20~\textmu s & 1~\textmu s & Measurement \\
    \begin{minipage}[t]{0.22\textwidth} 
    \ \\
    Energy resolution\\
    \ \\
    \end{minipage}
    &
    \begin{minipage}[t]{0.22\textwidth} 
    \ \\
    $\leq$19\% @60~keV \\
    \ \\
    \end{minipage}
    &
    \begin{minipage}[t]{0.23\textwidth} 
    GRD-A: 16.54\% @60~keV\\
    GRD-B: 16.53\% @60~keV\\
    GRD-C: 16.46\% @60~keV\\
    \end{minipage}
    & 
    \begin{minipage}[t]{0.22\textwidth} 
    \ \\
    Measurement \\
    \ \\
    \end{minipage} \\
    Burst observation rate & $>$90~/year & $>$100~/year & Simulation $\&$ Measurement \\
    GRB localization error & $<$5~degrees$^{\dagger}$ & $<$5~degrees$^{*}$ & Simulation $\&$ measurement \\
  \noalign{\smallskip}\hline
\end{tabular}
\end{center}
\tablecomments{0.86\textwidth}{$^{\dagger}$ Fluence $>1\times10^{-6}$~erg~cm\textsuperscript{-2} @1--1000~keV, 1~s.}
\tablecomments{0.86\textwidth}{$^{*}$ $1\sigma$ statistical error, Fluence $>1\times10^{-6}$~erg~cm\textsuperscript{-2} @1--1000~keV, 1~s. GRB spectral parameters: $\alpha = -1.9$,~$\beta = -3.7$,~$E_{peak} = 70$~keV.}
\end{table*}

The GRM subsystem consists of three primary units: the Gamma-Ray Detector (GRD), serving as the main detector for X-ray and soft gamma-ray spectral observations and GRB triggering; the GRM Particle Monitor (GPM), designed to monitor charged particle intensities and provide early warnings for regions like the SAA; and the GRM Electronics Box (GEB), which processes scientific data and manages command transmission and reception. Additionally, each GRD unit incorporates a GRM Calibration Detector (GCD) at the crystal edge for in-orbit gain monitoring and calibration. The physical configuration of GRM aboard SVOM is illustrated in Figure~\ref{Fig1}, with performance specifications and measured characteristics detailed in Table~\ref{Tab1}.

\begin{figure}[h] 
   \centering
   \includegraphics[width=8.0cm, angle=0]{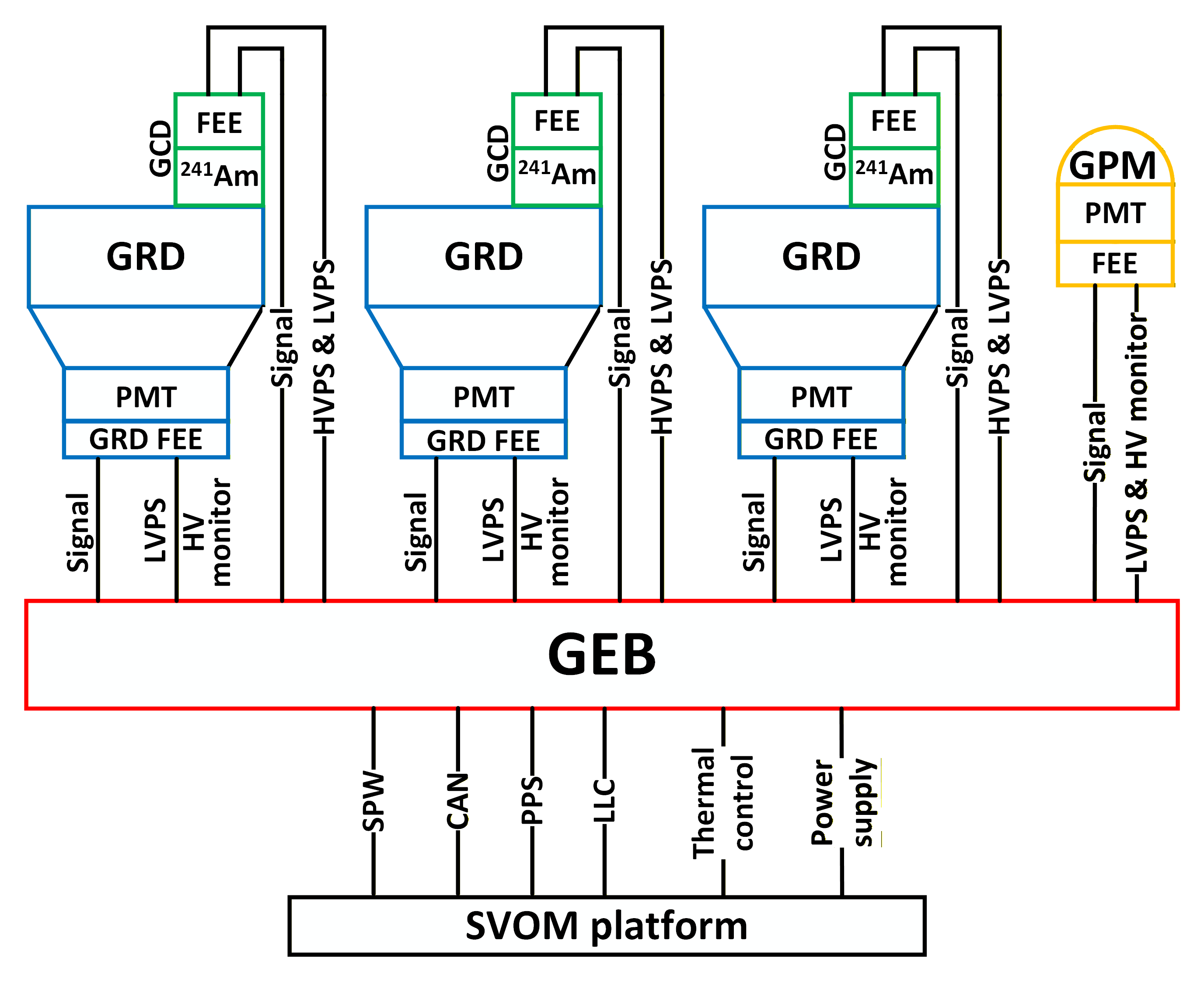}
   \caption{GRM system composition and component hierarchy.} 
   \label{Fig2}
\end{figure}

The complete component hierarchy of GRM is illustrated in Figure~\ref{Fig2}, with the GEB serving as the central processing unit responsible for detector configuration, data acquisition, and interface management with the SVOM platform. 

\begin{figure*}[t]
   \centering
   \includegraphics[width=14.0cm, angle=0]{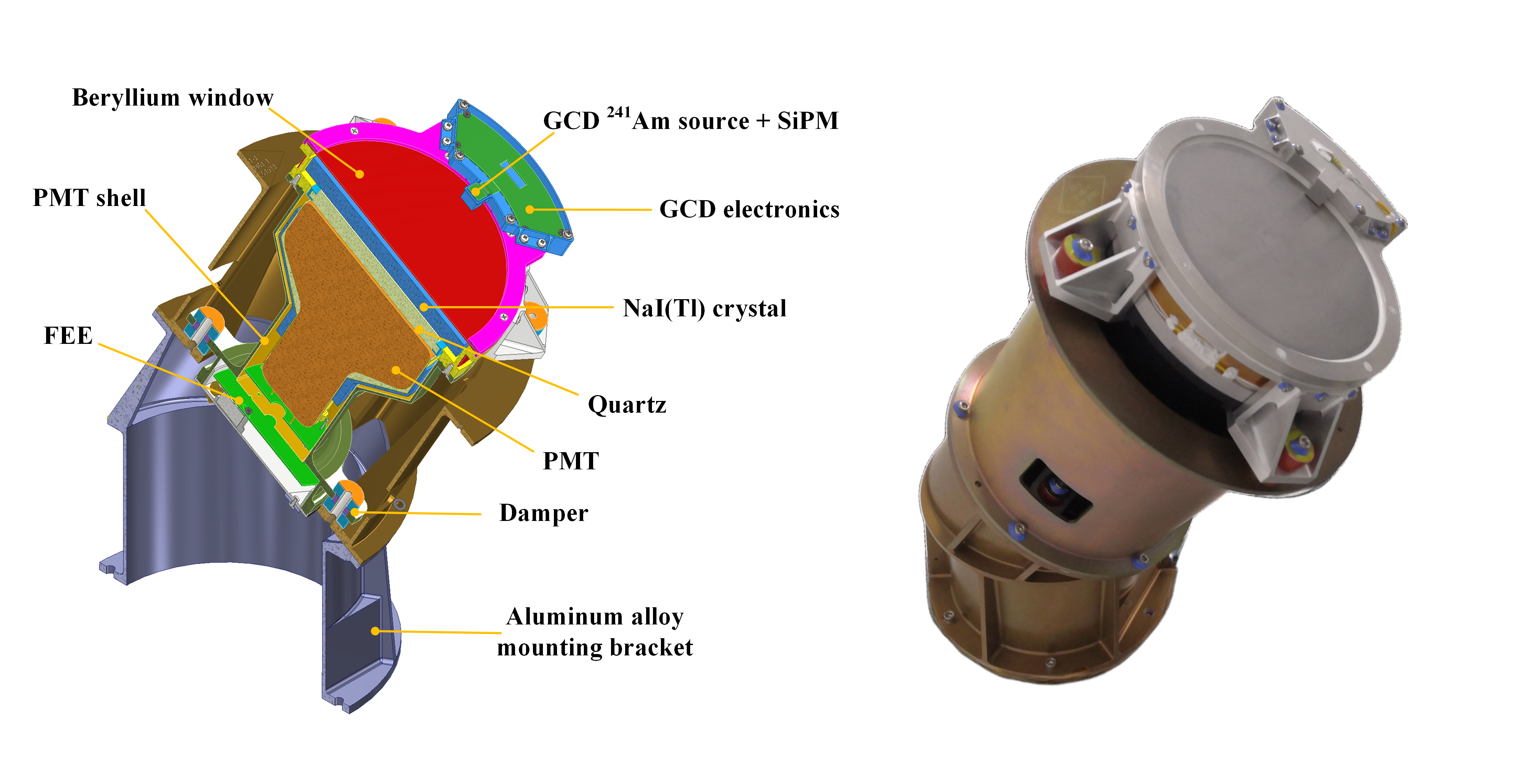}
   \caption{Design of the GRD assembly. Left: Exploded schematic view showing key components including a scintillator, photomultiplier tube (PMT), and front-end electronics (FEE). Right: Fully assembled flight model of a GRD unit, illustrating the final mechanical configuration used aboard the SVOM satellite.} 
   \label{Fig3}
\end{figure*}

The GEB performs two primary functions: first, it collects scientific data from the three GRD units, three GCD calibration detectors, and the GPM particle monitor, while simultaneously providing low-voltage power supply (LVPS), high-voltage power supply (HVPS), and high-voltage monitoring for these detectors; second, it manages communication interfaces with the satellite platform, utilizing SpaceWire (SPW) bus for scientific data transmission and Controller Area Network (CAN) bus for telemetry and command exchange.

The platform provides critical timing and control signals to GRM, including the Pulse Per Second (PPS) signal for precise time synchronization and the Low-Level Compatible (LLC) signal for system reset operations. Additionally, the platform supplies primary power and thermal control to the GRM subsystem through the GEB interface.

\section{Design}
\label{sect:instrument}

\subsection{Instruments}
\subsubsection{Gamma-Ray Detector}
\label{subsubsect:grd}
The GRD unit employs a NaI(Tl) scintillator to monitor hard X-ray and soft gamma-ray emissions in the 15--5000~keV energy range across a large portion of the sky. GRM incorporates three identical GRD units, each mounted to the payload instrument module (PIM) structure via a dedicated bracket. Similar to Fermi/GBM, the design omits collimators and backside anti-coincidence shielding. The three GRD units are oriented with elevation angles of 30° relative to the symmetry axis and azimuth angles spaced 120° apart, enabling coarse localization within a wide FoV—a capability particularly valuable for follow-up observations of gravitational wave associated events.

Each GRD unit comprises a Beryllium entrance window, a NaI(Tl) scintillator assembly, a photomultiplier tube (PMT), a shock absorption layer, magnetic shielding, front-end electronics (FEE), protective housing, and eight damping elements. FEE integrates amplifiers, a high-voltage module, a high-voltage control driver, and a high-voltage measurement circuit. A schematic diagram of the GRD assembly is provided in Figure~\ref{Fig3}.

To cover the full energy range given the limited dynamic range of FEE, each GRD utilizes two signal chains with different amplification factors: a high-gain channel for low-energy photons and a low-gain channel for high-energy events. The high-gain channel is linear over approximately 100–3500 ADC channels, and the low-gain channel also extends up to $\sim$3500 ADC channels. During calibration, the actual lower threshold of the high-gain channel was found to be below 100 ADC channels, corresponding to energies of several keV. The gain ratio between the two channels is approximately 9.5:1, with an overlapping energy response between 150~keV and 550~keV.

\subsubsection{GRM Calibration Detector}
\label{subsubsect:gcd}

The GRM Calibration Detectors (GCDs) are designed to monitor the gain stability of the GRD units and provide in-orbit calibration references. Each GRD is equipped with one GCD module positioned above its entrance window at the edge of the detector's FoV. The mechanical integration of the GCD on top of a GRD unit is illustrated in Figure~\ref{Fig3}.

Each GCD comprises an \textsuperscript{241}Am radioactive source, a Silicon Photomultiplier (SiPM) device, and associated readout electronics. The \textsuperscript{241}Am source emits $\sim$5~MeV alpha particles through alpha decay, accompanied by subsequent gamma-ray emissions. The SiPM detects the alpha particles, while the GRD registers the corresponding gamma-rays. By utilizing coincidence signals between the GCD and GRD, gamma photons from \textsuperscript{241}Am can be reliably identified for GRD onboard calibration purposes. The \textsuperscript{241}Am source has a total alpha activity of approximately 200~Bq, with the GRD detecting approximately 10\% of the gamma rays emitted by the source.

\begin{figure}[h]
   \centering
   \includegraphics[width=9.0cm, angle=0]{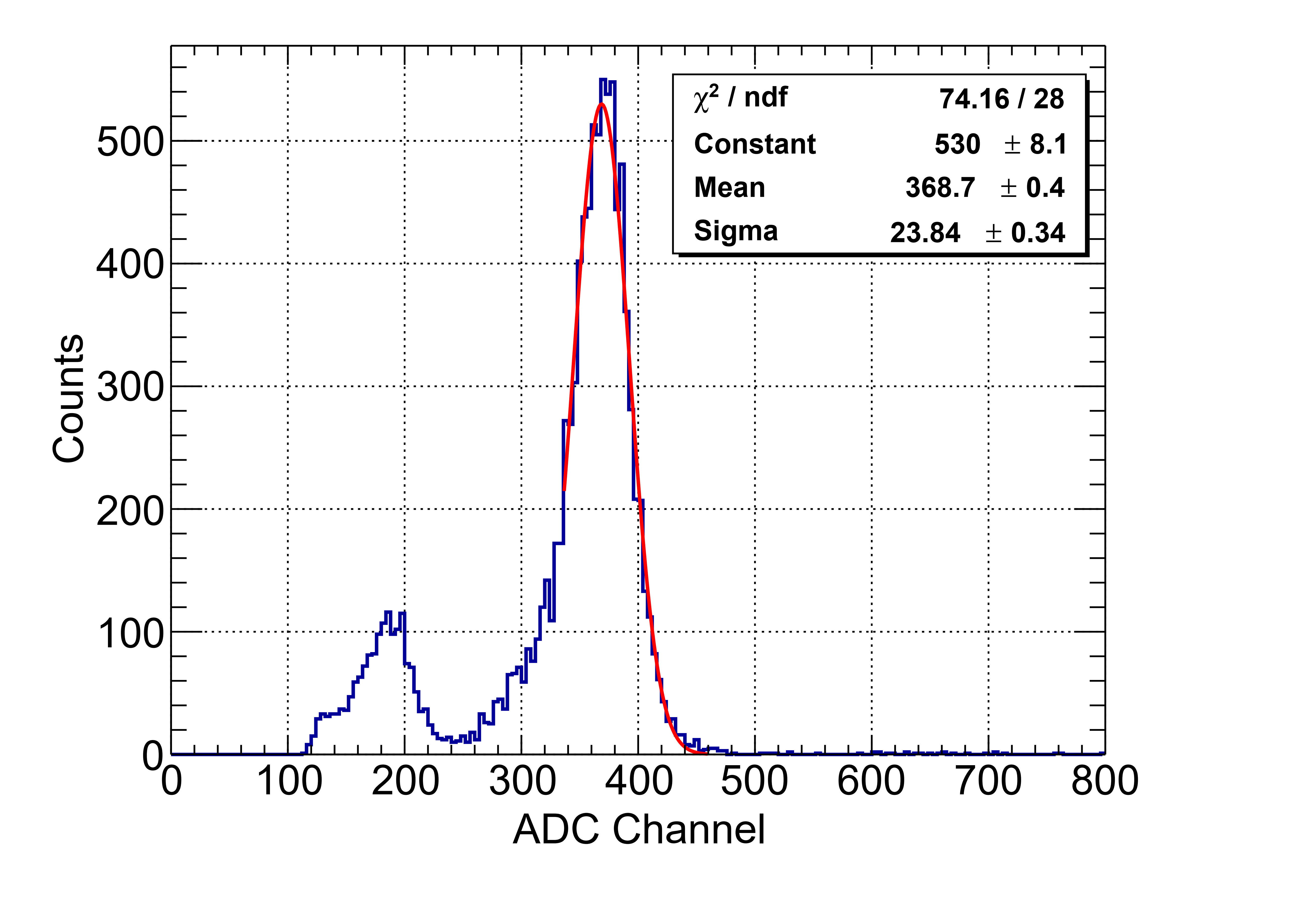}
   \caption{A typical \textsuperscript{241}Am spectrum measured with a GRD detector unit. The prominent photopeak on the right demonstrates the detector's energy response to the 59.5~keV emissions from the GCD, while the feature at lower energies includes both low-energy photoelectric peaks and characteristic escape peak from the detector material.}
   \label{fig-gcd-coincidence}
\end{figure}

For GRD gain monitoring, the prominent energy line at 59.5~keV from \textsuperscript{241}Am is utilized for low-energy calibration, while the in-orbit 511~keV annihilation background line provides a reference for higher energy calibration. With typical count rates of tens of gamma photons per second in the GRD, integration time intervals of 1--5 minutes are required to accumulate sufficient statistics for the calibration spectrum, and approximately 1 hour for the 511~keV line. Figure~\ref{fig-gcd-coincidence} shows a typical calibration spectrum measured by the GRD detector unit. The peak on the right of the spectrum corresponds to the 59.5~keV full-energy peak, achieving an energy resolution of better than 17\%. The left peak complex contains multiple low-energy full-energy peaks along with the escape peak.

To mitigate temperature-dependent gain variations in the SiPM, the power supply chain incorporates components with negative temperature coefficients that automatically compensate for these effects.

The mechanical design of the GCD minimizes the obstruction of the GRD's effective area. The \textsuperscript{241}Am source and SiPM are mounted on a dedicated section of the PCB, enclosed by a metallic frame that forms a light-tight darkroom. This enclosure is filled with a dark sealant to prevent external photon leakage and enhance mechanical stability, ensuring both optical isolation and shock resistance for safe operation of the radioactive source.

\subsubsection{GRM Charged Particle Monitor}
\label{subsubsect:gpm}

GPM, employing a BC-440M plastic scintillator coupled to a photomultiplier tube (PMT), is installed on the upper surface of the PIM. The cylindrical scintillator measures 10~mm in both diameter and height and is enclosed within a 2~mm thick aluminum housing. 

\begin{figure}[h]
   \centering
   \includegraphics[width=8.0cm, angle=0]{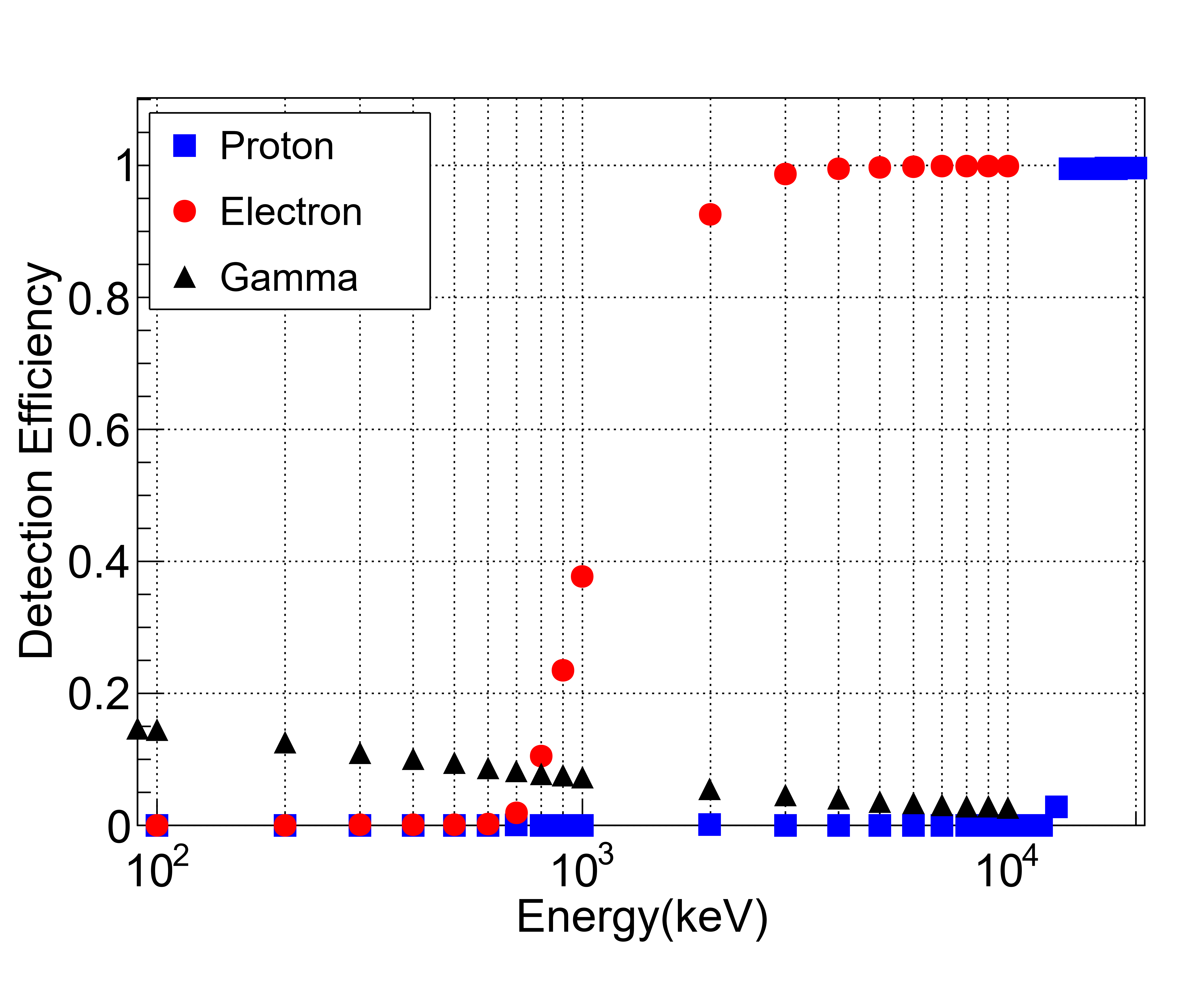}
   \caption{Simulated detection efficiency of GPM for electrons, protons, and gamma-ray photons as a function of incident particle energy. The distinct efficiency curves demonstrate the detector's capability to discriminate between charged particles and photons.}
   \label{fig-gpm-mc-effi}
\end{figure}

Based on simulation results (as shown in Figure~\ref{fig-gpm-mc-effi}), when the incident particle energy is 1~MeV, the detection efficiency of GPM for electrons is approximately 40\%, while for photons it is less than 10\%. With increasing energy, the detection efficiency for electrons becomes significantly higher than that for photons. Similarly, when the incident particle energy is approximately 10.5~MeV, the detection efficiency for protons approaches 100\%, while for photons it remains below 5\%. As the energy increases further, the detection efficiency for protons becomes substantially higher than that for photons. Therefore, GPM can effectively distinguish between high-energy charged particle bursts and GRBs. In the actual GPM detector configuration, the energy thresholds for high-energy electrons and protons are set at 1.3~MeV and 20~MeV, respectively, which enhances the detection of GRB photons while reducing false triggers from high-energy charged particles.

\begin{figure*}[t]
   \centering
   \includegraphics[width=13.0cm, angle=0]{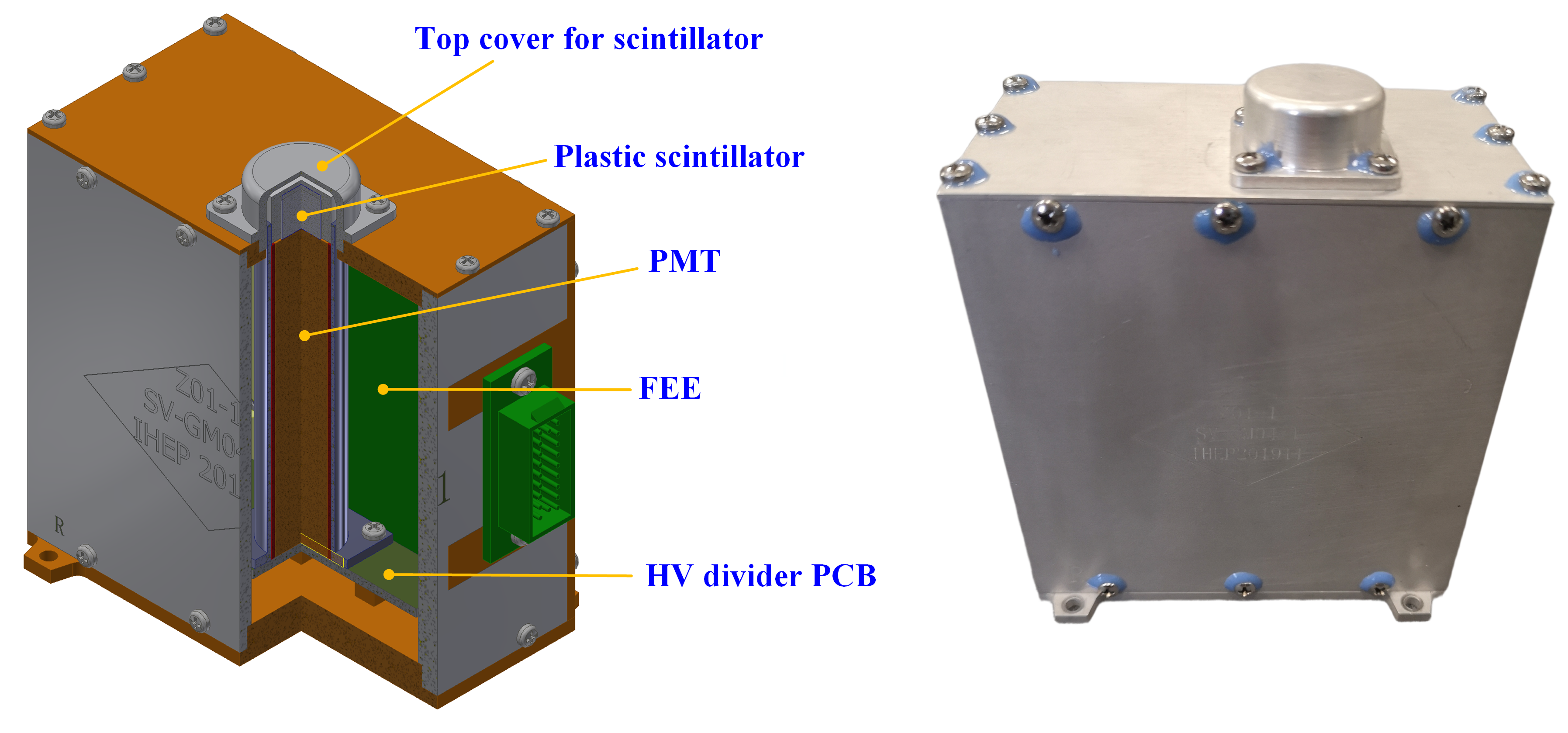}
   \caption{Design of the GPM assembly. Left: Exploded schematic view detailing the internal configuration, shielding structure, and detector components. Right: Fully assembled flight unit, demonstrating the compact mechanical design as implemented on the SVOM satellite for charged particle monitoring.}
   \label{Fig4}
\end{figure*}

The primary function of GPM is to monitor the in-orbit charged particle fluxes and automatically control the high-voltage supplies to the GRD PMTs. When the measured count rate exceeds a programmable threshold, the GPM triggers the shutdown of GRD high voltages; conversely, when the count rate falls below the threshold, the high voltages are restored. This protection mechanism safeguards the GRD detectors during periods of elevated radiation, such as passages through the SAA region.

The GPM assembly consists of an encapsulated enclosure, scintillator cover, FEE, high-voltage module, PMT assembly, and scintillator element. The scintillator cover serves dual purposes: establishing the energy threshold for detectable particles while providing mechanical protection for both the scintillator and PMT. The complete GPM design is illustrated in Figure~\ref{Fig4}.

\begin{figure*}[h]
   \centering
   \includegraphics[width=13.0cm, angle=0]{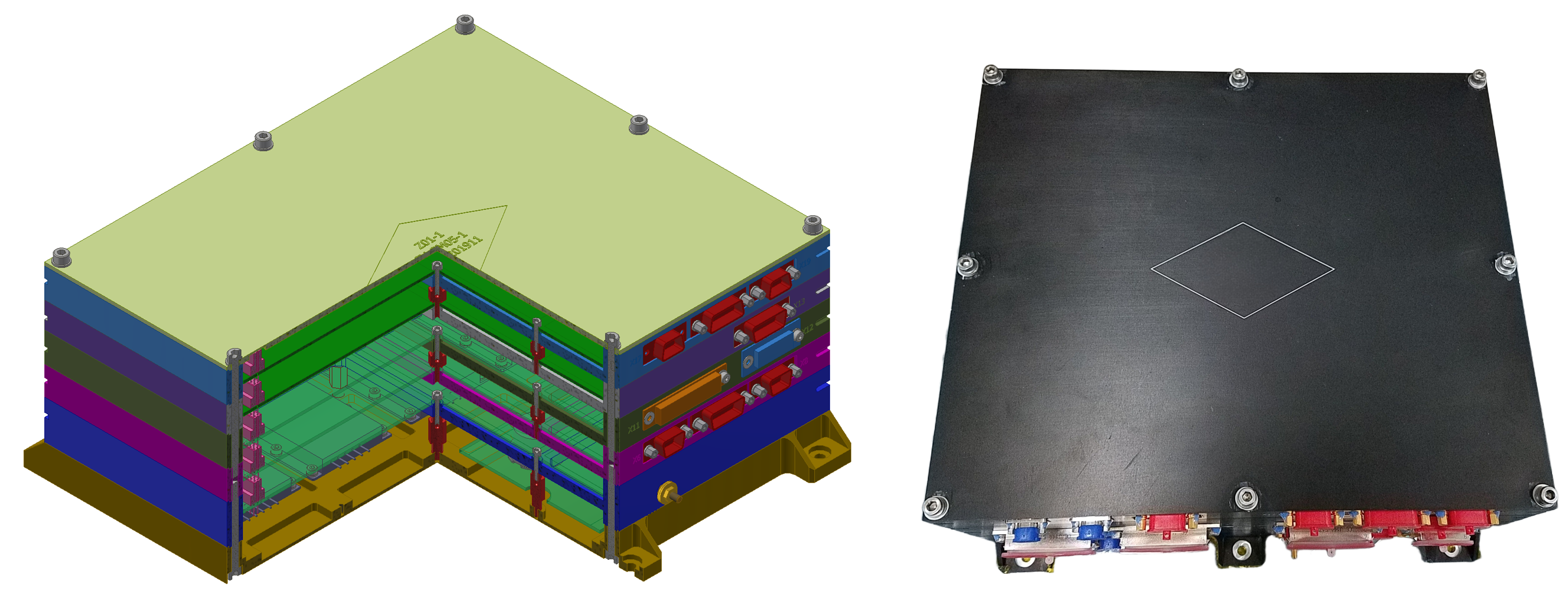}
   \caption{Design of the GEB assembly. Left: Exploded schematic view showing the internal organization of electronics. Right: Fully assembled flight unit featuring a black-anodized surface treatment for enhanced thermal and optical properties in the space environment.}
   \label{Fig5}
\end{figure*}

\subsubsection{GRM Electronics Box}
\label{subsubsect:geb}

GEB serves as the central processing and control unit for GRM, integrating multiple critical functions including data acquisition, data management, operational mode transitions, onboard trigger and localization algorithms, DC/DC power conversion, and communication interface management. GEB primarily consists of five electronic module layers: the Power Supply Unit (PSU), the primary data acquisition unit (CEU\_A\_N), the backup data acquisition unit (CEU\_A\_R), the primary data management unit (CEU\_D\_N), and the backup data management unit (CEU\_D\_R). The primary and backup modules feature identical designs and operate in a cold backup configuration. Each electronic module layer is composed of an aluminum frame, PCBs, and electronic components, with interlayer connectors distributed between all five circuit board layers. The complete GEB design configuration is presented in Figure~\ref{Fig5}.

\subsection{In-orbit Operation Modes}
\label{subsect:mode}

The GRM instrument operates through seven distinct operational modes, each designed to fulfill specific mission requirements and respond to varying space environment conditions.

\textbf{[Off Mode]}. In this mode, GRM is completely powered down. Transition from Off mode to other operational states is exclusively controlled by telecommand. This mode is typically activated during SVOM payload self-check procedures or when critical errors are detected in the GRM system.

\textbf{[Wait Mode]}. GRM low-voltage power is enabled while high-voltage systems remain powered off. The GEB awaits the LLC signal rising edge to initiate further operations in this standby state.

\textbf{[Configuration Mode]}. With low-voltage power active, GRM operates with default parameters. This mode allows configuration of high-voltage settings and other operational parameters via command. Transition to Background mode is exclusively command-initiated.

\textbf{[Background Mode]}. This is the nominal operational state when no GRBs or transient sources are detected. GRM collects rebinned spectral data at regular intervals and timing profiles across multiple energy bands. Event-by-event data are continuously recorded and transmitted to ground stations.

\textbf{[GRB1 (GRB-G) Mode]}. Activated automatically when GEB detects a significant count rate increase in GRD units at time $\mathrm{T_{0}}$, indicating a potential GRB. The instrument collects light-curve data from $\mathrm{T_{0}-2}$~min to $\mathrm{T_{0}+7}$~min for ground transmission. GRM automatically reverts to Background mode after $\mathrm{T_{0}+7}$~min. This mode can also be entered via ground command. Event-by-event data collection remains active throughout this mode.

\textbf{[GRB2 (GRB-E) Mode]}. Initiated either by ground command or upon trigger notification from the ECLAIRs instrument. Once activated, data collection follows the same temporal pattern as GRB1 mode ($\mathrm{T_{0}-2}$~min to $\mathrm{T_{0}+7}$~min) with automatic return to Background mode thereafter. Event-by-event data are preserved and transmitted.

\textbf{[SAA Mode]}. Activated by ground command when entering the SAA region (with tunable duration parameters). GRD high-voltage systems are disabled to protect the detectors from elevated radiation, while GPM continues monitoring charged particle fluxes. GRM exits SAA mode upon ground command when particle levels normalize, transitioning back to Background mode. SAA region parameters are ground-configurable.

\begin{figure}[t]
   \centering
   \includegraphics[width=8.0cm, angle=0]{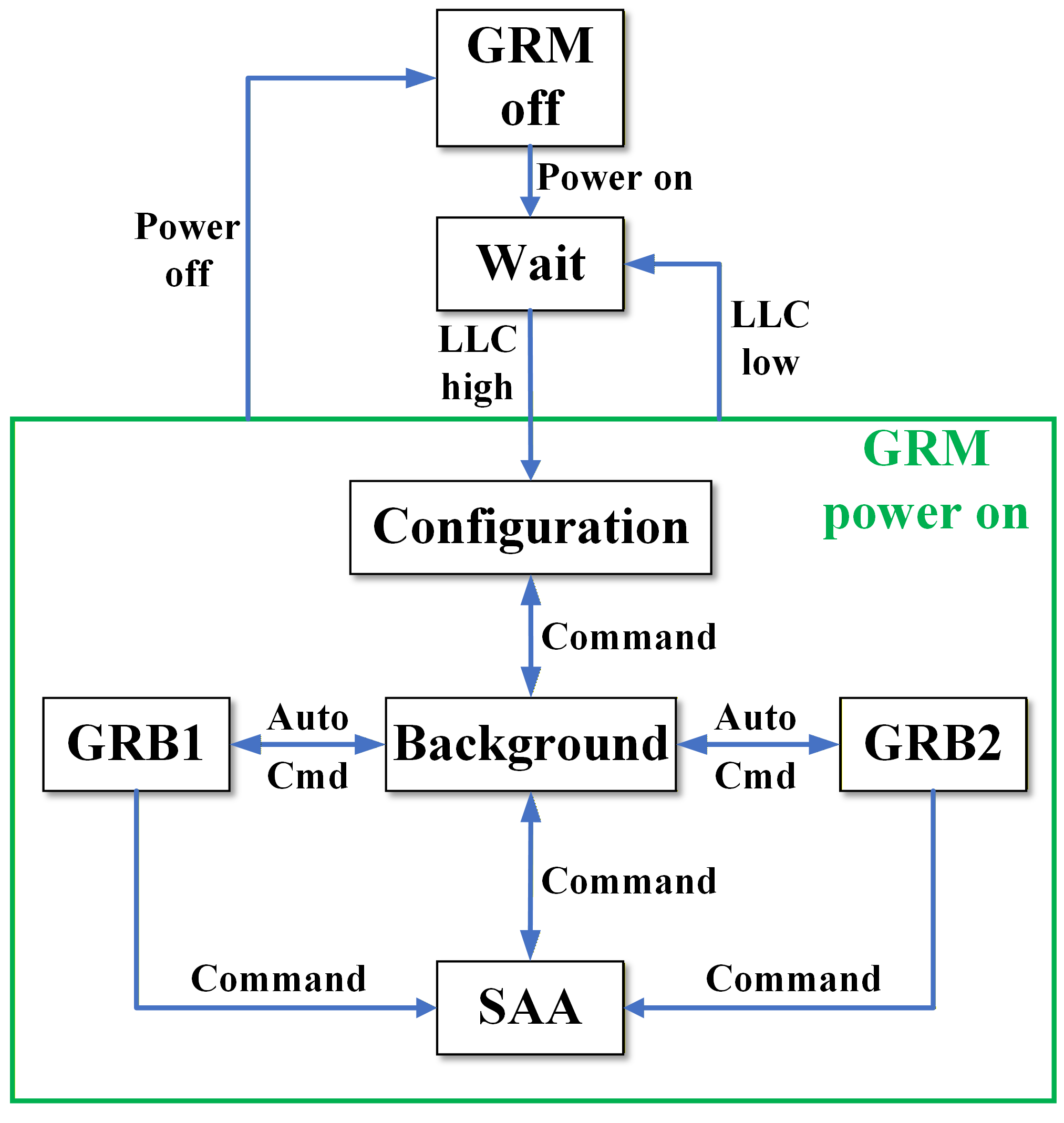}
   \caption{GRM operational mode transition diagram, illustrating the conditions and pathways between the seven defined operational states.}
   \label{Fig6}
\end{figure}

\textbf{[Mode Transition]}. GRM supports both automatic and command-initiated mode transitions. Automatic transitions occur based on internal instrument data and predefined triggers, while manual transitions are executed via platform commands. The complete mode transition logic is illustrated in Figure~\ref{Fig6}.

\subsection{On-board Trigger and Localization Algorithm}
\label{subsect:sw}

GRM employs a count rate trigger algorithm to detect GRBs by identifying significant increases in photon count rates. The trigger significance is calculated using Eq.~\eqref{eq1}:

\begin{equation}\label{eq1}
\text{Significance} = \frac{S - B}{\sqrt{B/T}},
\end{equation}
where $T$ is the duration of the burst window searched (in seconds), $B$ represents the time-normalized background count rate per second (without GRB signals) and $S$ denotes the source count rate per second (potentially containing both GRB signals and background events). GRM utilizes background estimates calculated over both short and long timescales: short-timescale backgrounds, calculated as averages immediately preceding the $S$ interval, are optimized for detecting short GRBs, while long-timescale backgrounds, accumulated over extended periods and fitted for precise characterization, reduce false trigger rates.

\begin{table*}[t]
\begin{center}
\caption{Parameters considered in the GRM on-board trigger algorithm.}\label{Tab2}
 \begin{tabular}{lc}
  \hline\noalign{\smallskip}
\textbf{Parameter} & \textbf{Values} \\
  \hline\noalign{\smallskip}
Energy ranges & 15--50 keV, 50--300 keV, 300--1000 keV, 1000--5000 keV \\ 
Timescales & 0.1 s, 1.0 s, 4.0 s \\
GRD combinations & GRD1+GRD2, GRD1+GRD3, GRD2+GRD3, GRD1+GRD2+GRD3 \\
  \noalign{\smallskip}\hline
\end{tabular}
\end{center}
\end{table*}

The trigger algorithm incorporates adjustable background accumulation periods and considers multiple detection parameters, including energy bands optimized for detector efficiency and GRB spectral characteristics, timescales matching typical GRB temporal features, and various GRD combinations for enhanced sensitivity. The specific parameters used in the trigger algorithm are detailed in Table~\ref{Tab2}. While the GRD combinations and timescales in Table~\ref{Tab2} are fixed in‑orbit, the energy ranges and the trigger thresholds (not listed) are adjustable via ground command. Their values were initially set based on pre‑launch simulations and are updated occasionally during operations based on observed trigger performance to optimize GRB trigger efficiency and reduce false triggers.

Based on the trigger algorithm, significances for the three GRDs across different energy band-timescale combinations are obtained. These are then compared against predefined thresholds to determine a trigger, according to the following logic: both Trigger Condition A and Trigger Condition B must be satisfied. Trigger Condition A requires that, in at least one energy band, the significances from at least two GRD units exceed their thresholds. Trigger Condition B requires that at least one combination of GRD units (as listed in Table~\ref{Tab2}) is triggered (significance exceeds the threshold).

For GRB localization, GRM utilizes the differential count rate responses across its three GRD units, which are mounted at 120° azimuthal intervals on the satellite's X-Y plane, each inclined at 60° to provide complementary sky coverage (Figure~\ref{Fig1}). Localization calculations are initiated only when all three GRDs are triggered simultaneously.

The on-board localization employs a lookup table approach using chi-squared minimization to match normalized measured count rates from the three GRDs with pre-simulated values. The celestial sphere is divided into 12 coarse regions, each further subdivided into 64 fine regions. The localization procedure involves: (a) identifying the coarse region; (b) refining within the corresponding fine region; and (c) applying chi-squared tests to determine the best match between measured and simulated count rates.

To accommodate different GRB spectral properties, three distinct localization tables (for soft, medium, and hard spectra) are employed based on the measured hardness ratio. Each table provides azimuth ($\phi$) and polar ($\theta$) angle coordinates and is stored in the GEB's EEPROM, with provisions for in-flight updates when necessary. Simulation-based validation confirms that GRM's on-board localization strategy meets the performance requirements outlined in Section~\ref{sect:mission}. A detailed analysis of GRM's trigger and localization performance is characterized in~\citep{grm_localization}.

\section{Ground Test and Validation}
\label{sect:groundtest}

\subsection{Ground-based Qualifications}
\label{subsect:qualification}

\begin{figure}[h]
   \centering
   \includegraphics[width=6.0cm, angle=0]{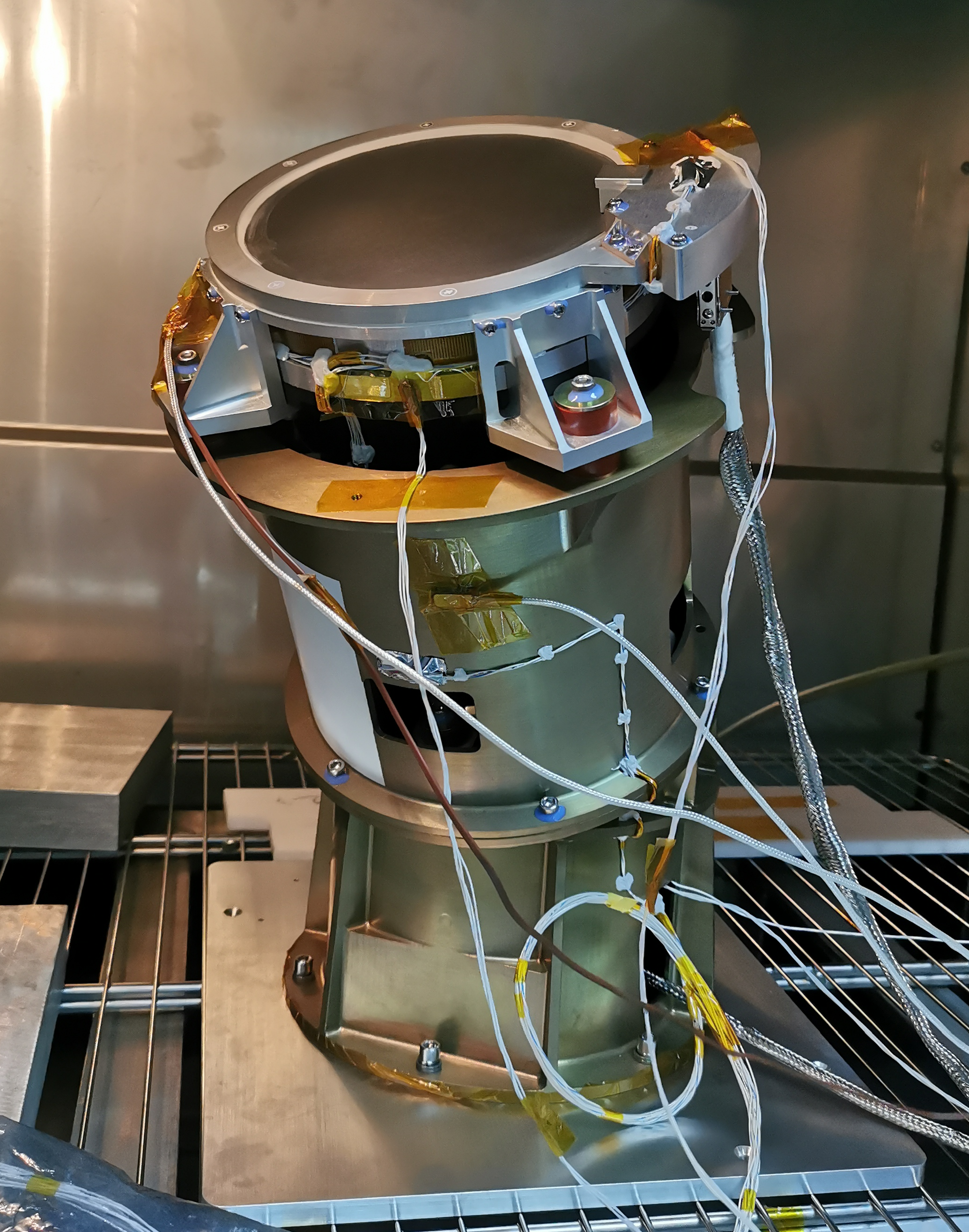}
   \caption{A GRD flight model unit undergoing thermal cycling testing. The unit is shown installed inside the thermal chamber, demonstrating the experimental setup used to verify the detector's performance under simulated space temperature environments ranging from $+5^{\circ}\mathrm{C}$ to $+35^{\circ}\mathrm{C}$ for 12.5 cycles, in accordance with the onboard product environmental testing requirements for the SVOM satellite.}
   \label{fig-grd-thermal}
\end{figure}

To verify the environmental adaptability and reliability of the GRM instruments, comprehensive qualification tests were conducted on both qualification and flight models prior to integration with the SVOM satellite platform. Following the satellite test program requirements, these tests included thermal cycling, thermal vacuum, vibration and shock, electromagnetic compatibility, and magnetic tests, all performed under strict quality control protocols. The test regimen comprehensively covered environmental conditions expected during both launch and in-orbit operations. Post-test review confirmed that the GRM subsystem successfully completed and passed all qualification tests. A photo of a GRD unit during the thermal cycling tests is shown in Figure~\ref{fig-grd-thermal}.

\begin{table*}[h]
\centering
\caption{GRD performance comparison before and after vibration and shock tests.}
\label{tab:perf_com_bf_vibshock}
\begin{tabular}{cccccccc}
  \hline
GRD & High Voltage & \multicolumn{3}{c}{\textsuperscript{241}Am Peak Position\textsuperscript{*}} & \multicolumn{3}{c}{Energy Resolution} \\
 & (V) & Before & After & Variation & Before & After & Variation \\
  \hline
GRD-A & 1175 & 369.90 & 372.30 & +0.65\% & 16.40\% & 16.51\% & +0.67\% \\
GRD-B & 1205 & 386.10 & 376.30 & -2.54\% & 16.55\% & 16.44\% & -0.66\% \\
GRD-C & 1185 & 371.30 & 379.80 & +2.29\% & 16.82\% & 16.80\% & -0.12\% \\
  \hline
\end{tabular}
\tablecomments{0.86\textwidth}{\textsuperscript{*}The 59.5~keV photon peak position of \textsuperscript{241}Am is given in ADC channels.}
\end{table*}

\begin{table*}[h]
\begin{center}
\caption{GRD performance comparison before and after thermal tests.}
\label{tab:perf_com_bf_thermal}
\begin{tabular}{cccc}
  \hline\noalign{\smallskip}
 & \textsuperscript{241}Am Peak position$^{*}$ & Energy resolution & High gain and low gain ratio \\
   \hline\noalign{\smallskip}
Before test & 384.00 & 17.76\% & 9.57 \\
After test & 373.00 & 17.73\% & 9.55 \\
Variation ratio & -2.86\% & -0.17\% & -0.21\% \\
  \noalign{\smallskip}\hline
\end{tabular}
\end{center}
\tablecomments{0.86\textwidth}{\textsuperscript{*}The 59.5~keV photon peak position of \textsuperscript{241}Am is given in ADC channels.}
\end{table*}

GRD performance was characterized using a \textsuperscript{241}Am radioactive source before and after each environmental test, with comparative analyses performed to verify the absence of significant performance degradation. Table~\ref{tab:perf_com_bf_vibshock} presents the performance comparison for the three GRD units before and after vibration and shock tests, while Table~\ref{tab:perf_com_bf_thermal} shows the results for thermal tests (thermal cycling and thermal vacuum) conducted on another GRD unit. The analysis demonstrates that variations in both peak positions and energy resolutions remain below 5\%, meeting all specified requirements.

\subsection{On-ground Test and Calibration}
\label{subsect:performance}

Following the ground calibration plan, comprehensive calibration experiments were conducted on both GRM qualification and flight models using hard X-ray beam facilities and multiple radioactive sources to characterize the instrument's detection performance. The energy response and detection efficiency were systematically measured across various operational configurations, including different high-voltage settings, discrimination thresholds, incident angles, temperature conditions, and beam positions.

\subsubsection{Hard X-ray Beam Tests}

The GRM hard X-ray beam calibration was conducted at the National Institute of Metrology's Hard X-ray Calibration Facility (HXCF). The facility provides monochromatic X-ray beams through two primary systems: double-crystal monochromators utilizing Si-220 crystals (35--80~keV) and Si-551 crystals (80--160~keV), and single-crystal monochromators employing LiF-220 or LiF-200 crystals (7--40~keV). A lead-shielded collimation system with precisely machined apertures (primary: 9~mm; secondary: 6~mm incident/5~mm exit) ensures well-defined beam geometry.

\begin{figure}[h]
   \centering
   \includegraphics[width=8.0cm, angle=0]{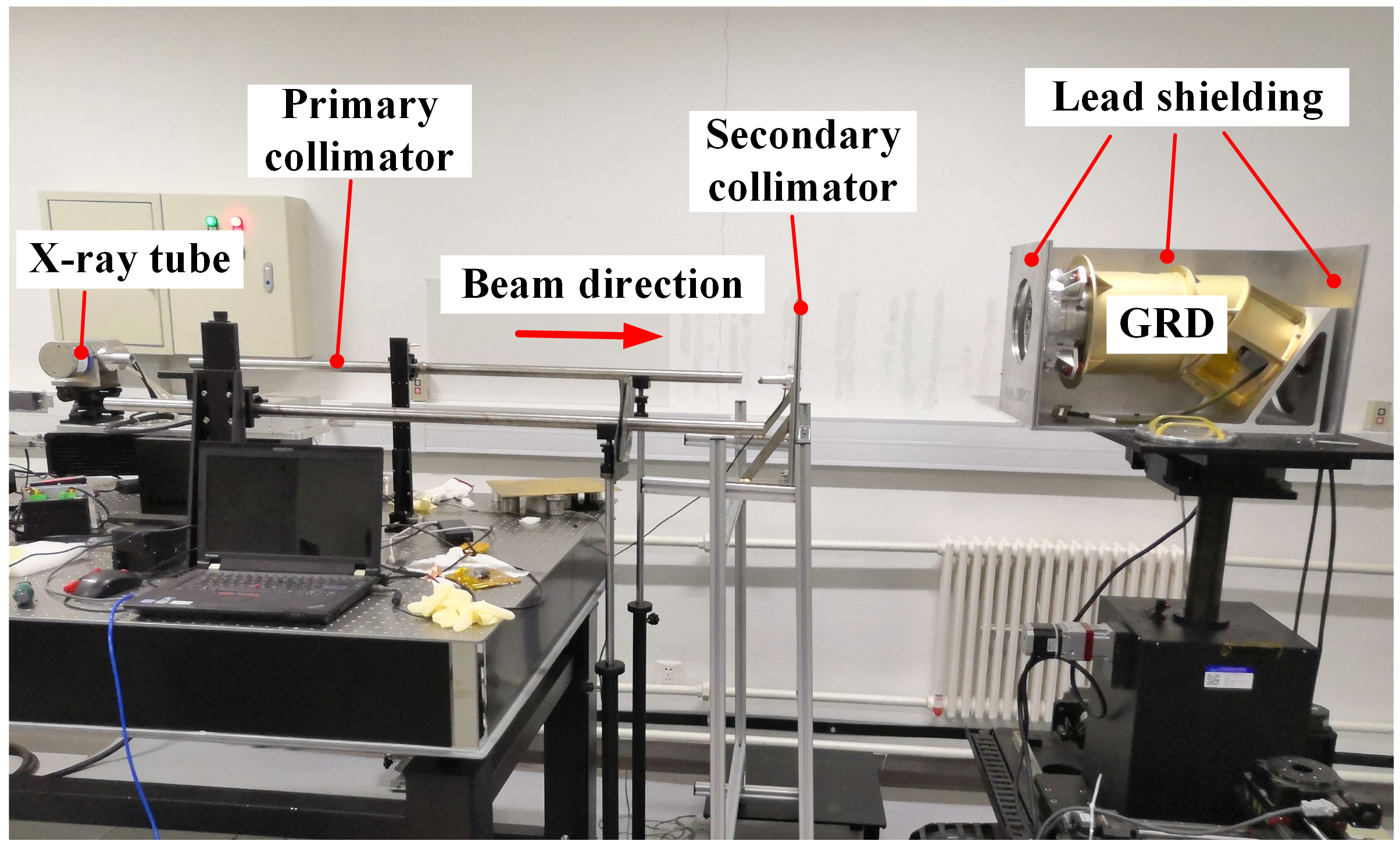}
   \caption{Experimental setup for GRM detector calibration at the Hard X-ray Calibration Facility (HXCF).}
   \label{Fig7}
\end{figure}

During calibration, each GRD detector was horizontally mounted on a translation stage within a shielding box, with the front surface exposed to the beam and positioned near the collimator exit. A high-purity germanium (HPGe) reference detector was placed alongside for independent beam monitoring, enabling comprehensive energy response characterization across the specified energy ranges. The complete experimental configuration at HXCF is shown in Figure~\ref{Fig7}.

\subsubsection{Radioactive Source Tests}

GRDs operate over an energy range of 15~keV to 5~MeV. To calibrate the high-energy response, multiple radioactive sources were employed to cover the entire detectable range and establish an accurate energy-channel (E-C) relationship. The calibration sources include \textsuperscript{137}Cs (661.6~keV, 32.2~keV), \textsuperscript{113}Sn (391.7~keV), \textsuperscript{60}Co (1173.2~keV, 1332.5~keV), \textsuperscript{88}Y (14.9~keV, 898.0~keV, 1836.1~keV), \textsuperscript{57}Co (14.4~keV, 122.1~keV), \textsuperscript{22}Na (511.0~keV, 1274.5~keV), and \textsuperscript{241}Am (59.5~keV).

\begin{figure}[h]
   \centering
   \includegraphics[width=8.0cm, angle=0]{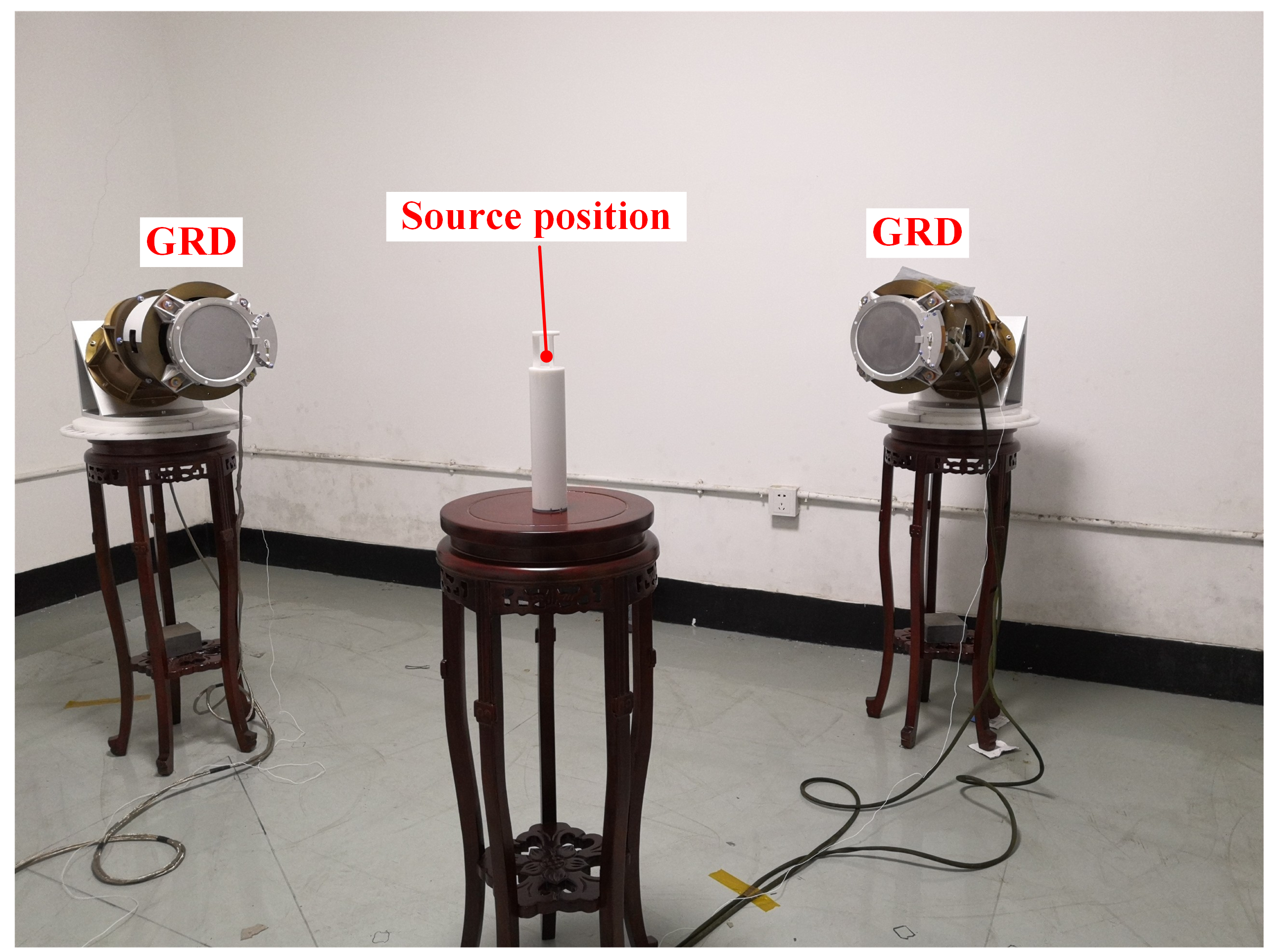}
   \caption{Experimental setup for GRM detector calibration using radioactive sources.}
   \label{Fig8}
\end{figure}

Utilizing the three data acquisition channels of GEB, three GRD units were calibrated simultaneously. To minimize detector non-uniformity effects and approximate parallel beam incidence, radioactive sources were positioned 3~m from the detector surfaces based on prior simulations. The three detector units were adequately spaced to reduce scattered photon interference. The complete source-based calibration setup is shown in Figure~\ref{Fig8}.

\subsubsection{Calibration Results}

\begin{figure}[h]
   \centering
   \includegraphics[width=8.0cm, angle=0]{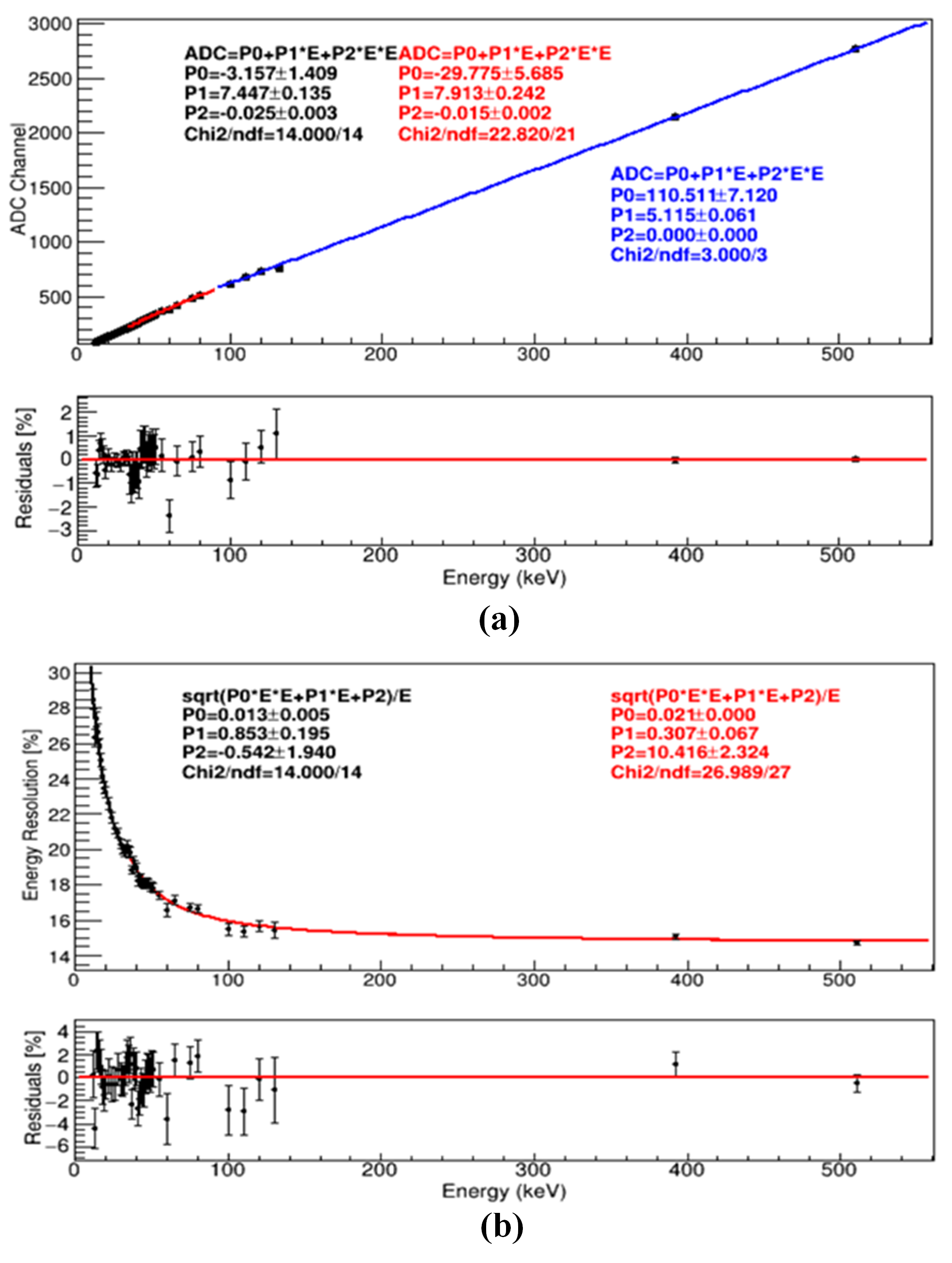}
   \caption{Preliminary calibration results for GRM. (a) Energy-channel relationship of GRD-A high-gain channel. (b) Energy resolution of GRD-A high-gain channel.}
   \label{Fig9}
\end{figure}

During calibration tests, the environmental temperature was maintained at $17 \pm 2\,^{\circ}\mathrm{C}$ to minimize temperature-induced variations. Combined analysis of HXCF beam and radioactive source calibration data provides a comprehensive detector performance characterization. Figure~\ref{Fig9} presents preliminary results for GRD-A's high-gain channel, showing the energy-channel (E-C) relationship and energy resolution. The E-C relationship was fitted in three energy segments while the energy resolution was fitted in two segments. Results indicate good energy linearity, with both energy range and resolution meeting specifications as indicated in Table~\ref{Tab1}. A detailed presentation of the complete ground calibration methodology and results will be provided in a dedicated GRM calibration paper currently in preparation.

\begin{figure}[h]
   \centering
   \includegraphics[width=7.0cm, angle=0]{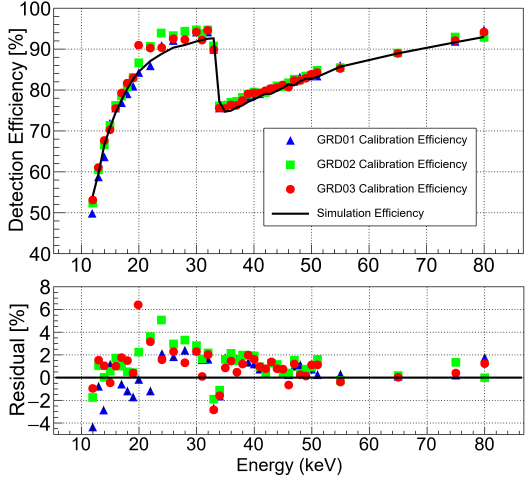}
   \caption{Comparison of measured and simulated detection efficiencies for the three GRD units across the 10--80~keV energy range. Experimental data were obtained through calibration measurements at the HXCF beamline, while simulation results were generated using the Geant4-based GRM mass model. The generally good agreement between measurement and simulation validates the accuracy of the detector response modeling, with deviations remaining within the 7\% requirement.}
   \label{fig-grd-efficiency}
\end{figure}

During the HXCF beam tests, by comparing the counting rates of a GRD unit and the HPGe reference detector, we can calculate the detection efficiency of the GRD. Due to the impact of beam flux stability on the calibration results of GRD's detection efficiency, the tests used comparative measurements from the HPGe detector to monitor the count rate variations of the beam before and after the GRD test. If the variation between two consecutive HPGe measurements exceeded 3\%, the beam was considered unstable, necessitating a recalibration. The detection efficiency measurement results of the three GRD units are shown in Figure~\ref{fig-grd-efficiency}, along with the simulation result. It indicates that the simulation result is generally consistent with the calibrated detection efficiencies for the three GRD units  across the 10--80~keV energy range, demonstrating that the Geant4 simulation mass model and physics interaction processes developed for GRM align well with the actual instrument design. However, a slight overestimation in the measured values is observed within the 20--30~keV energy range, which is likely due to instability in the beam flux during the tests. Despite this minor discrepancy, the deviation between the calibrated efficiencies and simulated values remains within 7\%, meeting the calibration accuracy requirements.

The uncertainties obtained during ground calibration provide a first-order estimate of the systematic uncertainties. The energy response residuals translate to a systematic uncertainty of a few percent in photon energy assignment across the energy range. For detection efficiency, the $<7\%$ deviation between measured and simulated values implies a similar level of systematic uncertainty in flux determination. These uncertainties may propagate into scientific products such as fluence and peak flux measurements at the level of a few percent, depending on the energy band and spectral shape.

To validate and refine the ground calibration results, GRM performs regular in-orbit calibrations using the onboard \textsuperscript{241}Am calibration sources and the 190 keV and 511 keV background spectral lines. Comparisons between in-orbit and ground calibration data allow for monitoring of gain drifts and correction of any residual offsets, ensuring the long-term stability and accuracy of the instrument response. The analysis of the GRM detector energy response is presented in~\citep{grm_energy_response}, and a comprehensive in-orbit calibration performance will be detailed in a dedicated paper currently in preparation.

\section{Preliminary In-orbit Performance}
\label{sect:in-orbit}

Since the launch of the SVOM satellite, GRM has maintained a nominal operational status. Throughout the commissioning phase, comprehensive parameter verification confirmed full instrument functionality consistent with ground-based design specifications. The onboard software demonstrates correct operation: GRM automatically disables high-voltage supplies upon entering the SAA area and restores them after exit; the instrument autonomously transitions between GRB1 (internally triggered) and GRB2 (externally triggered by ECLAIRs) modes, returning to background mode following burst episodes.

Preliminary in-orbit calibration confirms that the three GRD units maintain an energy response from several keV to several MeV (extending beyond 5 MeV), consistent with pre-launch calibration results. The energy resolution remains stable at 15\%--17\% across all units, with a temporal resolution better than 20~\textmu s. For bright GRBs (fluence $>1\times10^{-6}$~erg~$\cdot$~cm$^{-2}$ in 1--1000~keV, 1~s duration) within its FoV, GRM achieves a statistical localization accuracy better than 5$^\circ$~\citep{grm_localization}.

Between 26 June 2024 and 15 January 2025 (204 days of operation), GRM detected 78 high-energy transients, including 71 confirmed GRBs. This detection rate corresponds to an annual yield exceeding 100 GRBs, surpassing the design requirement of 90 GRBs per year and in good agreement with pre‑launch simulations, which predicted an annual detection rate of approximately 106~GRBs~\citep{he2025svom}.

\begin{figure}[h]
   \centering
   \includegraphics[width=8.0cm, angle=0]{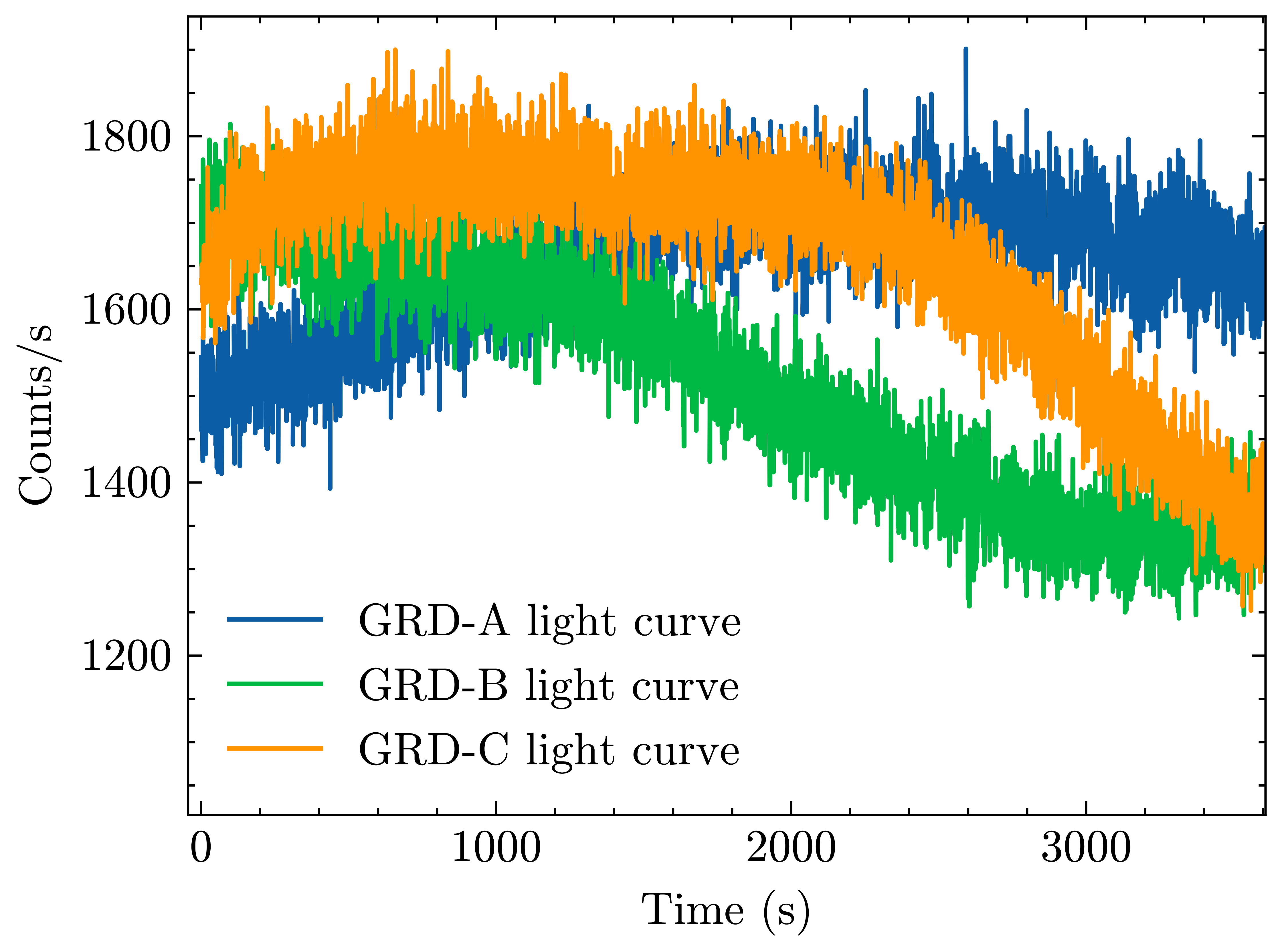}
   \caption{In-orbit background light curves for all three GRD units. The curves show one-hour observations in the 15--5000~keV energy band starting from 2025-10-01T22:44:00.}
   \label{Fig10}
\end{figure}

\begin{figure}[h]
   \centering
   \includegraphics[width=8.0cm, angle=0]{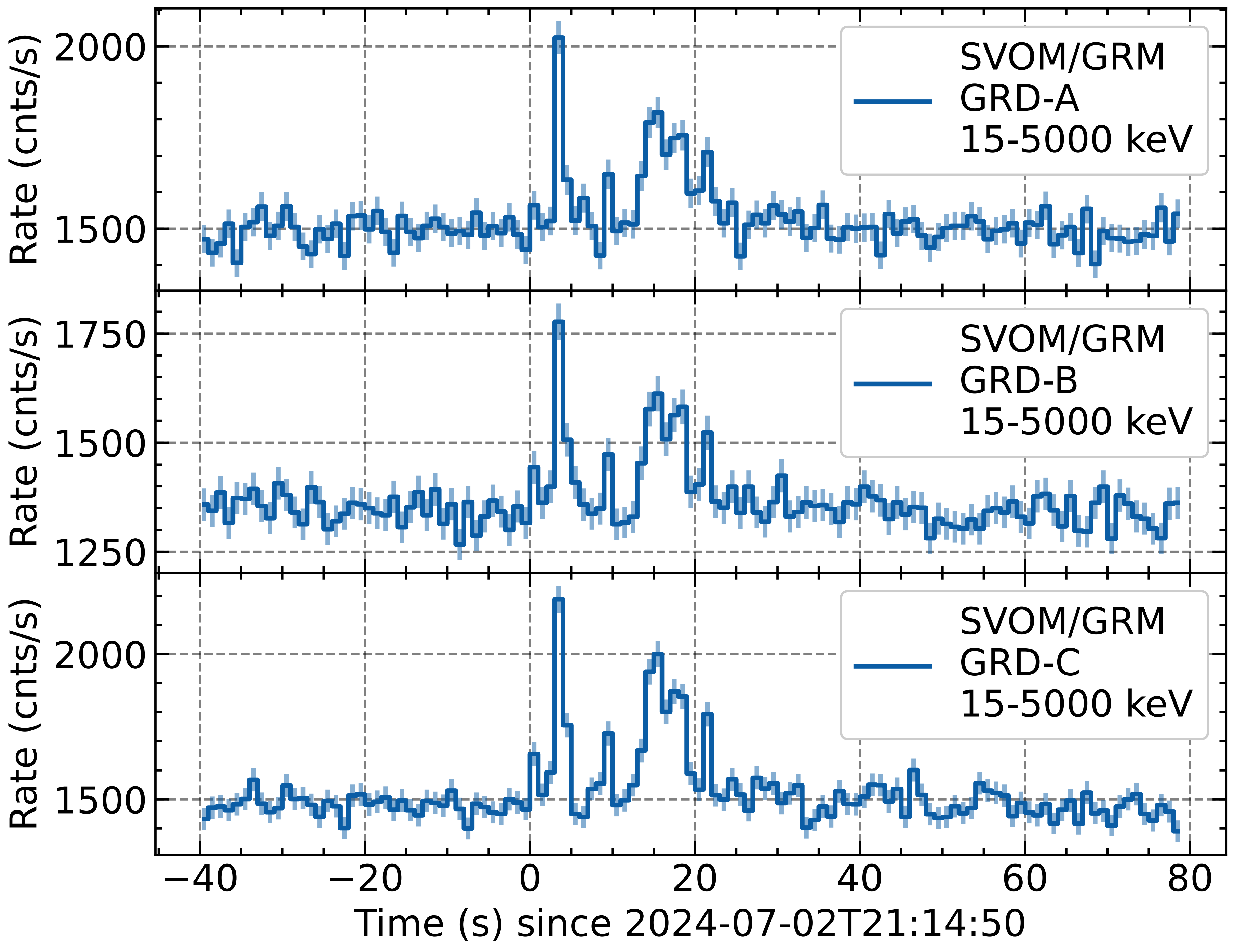}
   \caption{Light curves of the bright long-duration GRB 240702A observed by the three GRDs of SVOM/GRM. The emission exhibits a multi-peaked temporal structure.}
   \label{Fig:GRB240702A}
\end{figure}

Figure~\ref{Fig10} presents the in-orbit background light curves for all three GRD units over a one-hour period. The count rates fluctuate between approximately 1200--1800~counts/s, showing excellent agreement with Monte Carlo simulations \citep{he2020bk}, which predicted an average background rate of $\sim$1340~counts/s for typical orbital conditions. The observed background differences among the three units result from their distinct pointing directions and varying obstruction conditions, as expected from pre-launch analysis. Figure~\ref{Fig:GRB240702A} shows a bright GRB (GRB 240702A) detected by the GRM during the early in-orbit phase~\citep{svom1stgcn}, which was also detected by other instruments such as the GBM~\citep{2024GCN.36802....1F}. Further analysis of GRM's in-orbit performance is ongoing, with detailed publications in preparation.

\section{Summary}
\label{sect:summary}

The GRM instrument onboard the SVOM satellite has been operating successfully in orbit for more than 1.5 years, demonstrating excellent stability and reliability. Throughout this period, the instrument has maintained consistent performance while detecting numerous high-energy transient events, particularly GRBs, contributing significantly to time-domain astronomy and multi-messenger astrophysics.

The comprehensive in-orbit verification confirms that all key performance parameters—including the energy range (extending from several keV to beyond 5 MeV), energy resolution (15\%--17\%), temporal resolution (better than 20~\textmu s), and background characteristics—align closely with pre-launch predictions based on extensive ground calibration campaigns and Monte Carlo simulations. All functional and performance requirements specified in the mission design have been successfully fulfilled, validating the instrument's robust design and implementation. The scientific productivity of GRM is evidenced by the detection of over 100 GRBs annually, exceeding the mission requirement of 90 GRBs per year. These observations have enabled valuable insights into GRB spectral properties, temporal characteristics, and population statistics. The instrument's capability to provide coarse localization and autonomous mode transitions has proven essential for rapid follow-up observations and multi-wavelength coordination.

The continued operation of GRM promises further contributions to high-energy astrophysics. Ongoing data analysis is yielding new insights into GRB physics and other transient phenomena~\citep{2025ApJ...995L..16T}. A series of publications detailing the instrument performance, calibration methodology, and scientific discoveries are in preparation and will be submitted to peer-reviewed journals in the coming months.

\begin{acknowledgements}
The Space-based multi-band Variable Objects Monitor (SVOM) is a joint Chinese-French mission led by the Chinese National Space Administration (CNSA), the French Space Agency (CNES), and the Chinese Academy of Sciences (CAS). We gratefully acknowledge the unwavering support of NSSC, IAMCAS, XIOPM, NAOC, IHEP, CNES, CEA, and CNRS. The authors are grateful for the support from the National Key R$\&$D Program of China (Grant No. 2024YFA1611701 and 2024YFA1611700), 
the National Natural Science Foundation of China (Grant No. 
12494572, 
12273042, 
11961141013 and 
12333007
), the Special Exchange Program A of Chinese Academy of Sciences (Grant No.~2H2025000112), the Special Program for Enhancing Original Innovation Capability of Chinese Academy of Sciences (Grant No.~292024000260) and China's Space Origins Exploration Program. The published material is based on observations performed by the SVOM mission funded by the respective space agencies (CNSA and CNES).
\end{acknowledgements}

\appendix                  

\bibliographystyle{raa}
\bibliography{bibtex}

\label{lastpage}

\end{document}